\documentclass[twocolumn,prb,color,psfig,superscriptaddress]{revtex4}
\usepackage{graphicx}
\usepackage{epsfig}
\begin{document}

\title{Topological phase separation  in 2D
quantum lattice Bose-Hubbard system away from half-filling}
\author{A.S. Moskvin}
\affiliation{Ural State University, 620083, Ekaterinburg,  Russia}

\date{\today}
\begin{abstract}
We  suppose that the doping of the 2D
hard-core boson system away from half-filling may result in the formation of
multi-center topological inhomogeneity (defect) such as charge order (CO)
bubble domain(s) with Bose superfluid (BS) and extra bosons both localized in
domain wall(s), or a {\it topological} CO+BS {\it phase
separation}, rather than an uniform mixed CO+BS supersolid phase. Starting from
the classical model we predict the properties of the respective quantum system.
The long-wavelength behavior of the  system is believed to remind that of
granular superconductors, CDW materials,  Wigner crystals, and multi-skyrmion
system akin in a quantum Hall ferromagnetic state of a 2D electron gas.

To elucidate the role played by quantum effects and that of the lattice
discreteness we have addressed the simplest nanoscopic counterpart of the bubble
domain in a checkerboard CO phase of 2D hc-BH square lattice. It is shown that
the relative magnitude and symmetry of multi-component order parameter are
mainly determined by the sign of the $nn$ and $nnn$ transfer integrals. In
general, the topologically inhomogeneous phase of the hc-BH system away from
the half-filling can exhibit the signatures both of $s,d$, and $p$ symmetry of
the off-diagonal order.

\end{abstract}

\maketitle

\section{Introduction}
 The model of quantum lattice Bose gas has a long history and has been
suggested initially for conventional superconductors
\cite{Schafroth}, quantum crystals such as $^{4}$He where
superfluidity coexists with a crystalline
order.\cite{Matsuda,Fisher} Afterwards, the Bose-Hubbard (BH) model
has been studied as a model of the superconductor-insulator
transition in materials with  local bosons, bipolarons, or preformed
Cooper pairs.\cite{Kubo,RMP} Two-dimensional BH models have been
addressed as relevant to describe the superconducting films and
Josephson junction arrays. The most recent interest to the system of
hard-core bosons comes from the delightful results on Bose-Einstein
(BE) condensed atomic systems produced by trapping bosonic neutral
atoms in an optical lattice. \cite{Greiner}
 The progress in boson physics
generates an especial interest  around nonlinear topological excitations
(skyrmions, vortices) which play an increasingly significant role in physics.

One of the fundamental hot debated  problems in bosonic physics concerns the
evolution of the charge ordered (CO) ground state of 2D hard-core BH model
(hc-BH)  with a doping away from half-filling.
Numerous model studies steadily confirmed the emergence of "supersolid" phases
with simultaneous diagonal and off-diagonal long range order while Penrose and
Onsager \cite{Penrose} were the first showing as early as 1956 that supersolid
phases do not occur.
The most recent quantum Monte-Carlo  (QMC) simulations
\cite{Batrouni,Hebert,Schmid} found two significant features of the
2D Bose-Hubbard model with a
 screened Coulomb repulsion: the absence of supersolid
phase  at half-filling, and a  growing tendency to phase separation (CO+BS)
upon doping away from half-filling.  Moreover, Batrouni and Scalettar
\cite{Batrouni} studied quantum phase transitions in the ground state
of the 2D hard-core boson Hubbard Hamiltonian and have shown  numerically that,
contrary to the generally held belief, the most commonly discussed
"checkerboard" supersolid is thermodynamically unstable and phase separates
into solid and superfluid phases. The physics of the CO+BS phase separation in
Bose-Hubbard model is associated with a rapid increase of the energy of a
homogeneous CO state with doping away from half-filling due to a large
"pseudo-spin-flip" energy cost.
Hence, it appears to be
energetically more favorable to "extract" extra bosons (holes) from the CO
state and arrange them into finite clusters with a relatively small number of
particles. Such a droplet scenario is believed  to minimize the long-range
Coulomb repulsion.
  However, immediately there arise several questions. Whether a simple
mean-field approximation  (MFA)
and classical continuum model can predict such a behavior? What is the detailed
structure of the CO+BS phase separated state? What are the main factors
governing the essential low-energy and long-wavelength physics? Is it possible
to make use of simple toy models? Are there the analogies with fermion Hubbard model?
The behavior of the latter under doping away from half-filling is extensively studied in last decade in frames of stripe scenario which implies  
 that doping may proceed through the formation of stripes, or charged domain walls being specific topological solitons in N\'{e}el antiferromagnets.
 In particular,
Emery and Kivelson \cite{Emery} argued that  the
phase separation reflects a universal tendency of the correlated antiferromagnet to expel the doped holes. White and Scalapino \cite{Scalapino} showed that  the pure 2D t-J model, in the small-J/t regime  and with dopings near $x\sim 0.1$, has a striped ground state. In general, the topological phase separation is believed to be a generic property of 2D fermion Hubbard model.

In the paper we
present a topological scenario of CO+BS phase separation in 2D hc-BH model with
inter-site repulsion. The extra
bosons/holes doped to a checkerboard CO phase of 2D boson system are believed
to be confined in the ring-shaped domain wall of the skyrmion-like topological
defect which looks like a bubble domain in an easy-axis antiferromagnet. This
allows us to
propose the mechanism of 2D {\it topological} CO+BS {\it phase separation} when
the
doping of the bare checkerboard CO phase results in the formation of a
multi-center topological defect, which simplified pseudo-spin pattern looks
like  a system of bubble CO domains with Bose superfluid confined in  charged
ring-shaped domain walls.

The rest of the paper is organized as follows. In Sec.II we give a
short overview of the conventional hc-BH model in frames of a
pseudo-spin formalism and MFA description. In Sec.III we show that
the doping in 2D  hc-BH system can be accompanied by the formation
of a topological defect like a single or multi-center skyrmion. We
present the scenario of the essential low-energy physics  for the
topologically doped hc-BH system. In Sec.IV we address a simple
model of nanoscopic bubble-like domain in a checkerboard CO phase
for a discrete square lattice that allows us to demonstrate a subtle
microscopic structure of such a center with a multi-component order
parameter.

\section{Hard-core Bose-Hubbard model}
\subsection{Effective Hamiltonian. Pseudo-spin formalism}
The Hamiltonian of hard-core Bose gas on a lattice can be written in a standard
form as follows:
\begin{equation}
\smallskip
H_{BG}=-\sum\limits_{i>j}t_{ij}{\hat
P}(B_{i}^{\dagger}B_{j}+B_{j}^{\dagger}B_{i}){\hat P}
+\sum\limits_{i>j}V_{ij}N_{i}N_{j}-\mu \sum\limits_{i}N_{i},  \label{Bip}
\end{equation}
where ${\hat P}$ is the projection operator which removes double occupancy of
any site,  $N_{i}=B_{i}^{\dagger}B_{i}$, $\mu $  the chemical potential
determined from the condition of fixed full number of bosons $N_{l}=
\sum\limits_{i=1}^{N}\langle N_{i}\rangle $ or concentration $\;n=N_{l}/N\in
[0,1]$. The $t_{ij}$ denotes an effective transfer integral,  $V_{ij}$ is an
intersite interaction between the bosons. \smallskip Here $B^{\dagger}(B)$ are
the Pauli creation (annihilation) operators which are Bose-like commuting for
different sites $[B_{i},B_{j}^{\dagger}]=0,$ if $i\neq j,$  $[B_{i},B_{i}^{\dagger}]=1-2N_i$,
$N_i = B_{i}^{\dagger}B_{i}$; $N$ is a full number of sites. It is worth noting that near half-filling ($n\approx 1/2$) one might introduce the renormalization $N_i \rightarrow (N_i -1/2)$, or neutralizing background, that immediately provides the particle-hole symmetry.  

\smallskip The model of hard-core Bose-gas with an intersite repulsion is
equivalent to a system of spins $s=1/2$  exposed to an external magnetic field
in the $z$-direction. For the system with neutralizing background we arrive at an effective pseudo-spin Hamiltonian
\begin{equation}
H_{BG}=\sum_{i>j}J^{xy}_{ij}(s_{i}^{+}s_{j}^{-}+s_{j}^{+}s_{i}^{-})+\sum\limits
_
{i>j}
J^{z}_{ij}s_{i}^{z}s_{j}^{z}-\mu \sum\limits_{i}s_{i}^{z}, \label{spinBG}
\end{equation}
where $J^{xy}_{ij}=2t_{ij}$, $J^{z}_{ij}=V_{ij}$, $s^{-}= \frac{1}{\sqrt{2}}B_ , s^{+}=-\frac{1}{\sqrt{2}}
 B^{\dagger}, s^{z}=-\frac{1}{2}+B_{i}^{\dagger}B_{i}$,
$s^{\pm}=\mp \frac{1}{\sqrt{2}}(s^x \pm \imath s^y)$.

 In a linear approximation the Hamiltonian for the coupling with an
electromagnetic field reads as follows
\begin{equation}
\hat V_{int}=\sum _{i<j}\hat t_{ij}((\Phi _{j}-\Phi _{i})[{\hat {\bf
s}_{i}}\times {\hat {\bf s}_{j}}]_{z}-\frac{1}{2}(\Phi _{j}-\Phi
_{i})^{2}({\hat {\bf s}_{i}}\cdot {\hat {\bf s}_{j}}));
\end{equation}
\begin{eqnarray}
(\Phi _{j}-\Phi _{i})=-\frac{q}{\hbar c}\int _{{\vec R}_{i}}^{{\vec R}_{j}}{\vec A}({\vec r})d{\vec r},
\end{eqnarray}
where  ${\vec A}$ is the vector-potential, and integration runs over line binding the  $i$ and $j$ sites; 
\begin{equation}
({\hat {\bf s}_{i}}\cdot {\hat {\bf s}_{j}})=\frac{1}{2}(\hat
B_{i}^{\dagger}\hat B_{j}+\hat B_{i}\hat B_{j}^{\dagger}),\,
[{\hat {\bf s}_{i}}\times {\hat {\bf s}_{j}}]_{z}=\frac{i}{2}(\hat
B_{i}^{\dagger}\hat B_{j}-\hat B_{i}\hat B_{j}^{\dagger}),
\end{equation}
however, the pseudo-spins are assumed to lie in $xy$ plane.
Then the boson current density operator one may represent to be a sum of the
paramagnetic
\begin{eqnarray}
{\bf j}^{p}({\bf R}_{i})=\frac{q}{\hbar }\sum _{j}{\hat t}_{ij}{\bf
R}_{ij}[{\hat {\bf s}_{i}}\times {\hat {\bf s}_{j}}]_{z},
\end{eqnarray}
and diamagnetic
\begin{eqnarray}
{\bf j}^{d}({\bf R}_{i})=-\frac{q}{2\hbar}\sum _{j}\hat t_{ij}{\bf R}_{ij}
(\Phi _{j}-\Phi _{i})({\hat {\bf s}_{i}}\cdot {\hat {\bf s}_{j}})
\end{eqnarray}
terms, respectively.

\subsection{Uniform phases: mean-field approximation}
Below we make use of a conventional two-sublattice MFA approach.  For the
description of the pseudospin
ordering to be more physically clear one may introduce two vectors, the
ferromagnetic and antiferromagnetic ones:
$$
{\bf m}=\frac{1}{2s}(\langle{\bf s}_1 \rangle +\langle{\bf s}_2 \rangle);\,
{\bf l}=\frac{1}{2s}(\langle{\bf s}_1 \rangle -\langle{\bf s}_2
\rangle);\,\,{\bf m}^2 +{\bf l}^2 =1\, .
$$
where ${\bf m}\cdot {\bf l}=0$. For the plane where these vectors
lie one may introduce two-parametric angular description: $m_x =
\sin \alpha \cos\beta ,m_z = -\sin \alpha \sin\beta , l_x = \cos
\alpha \sin\beta , l_z = \cos \alpha \cos\beta $. The hard-core
boson system in a two-sublattice approximation is described by two
diagonal order parameters $l_z ,m_z$ and two complex off-diagonal
 order parameters $m_{\pm}=\mp \frac{1}{\sqrt{2}}(m_x \pm \imath m_y)$ and
$l_{\pm}=\mp \frac{1}{\sqrt{2}}(l_x \pm \imath l_y)$. The complex superfluid
order parameter $\Psi ({\bf r})=|\Psi ({\bf r})|\exp(-\imath\varphi ) $ is
determined by the in-plane components of ferromagnetic vector: $ \Psi ({\bf
r})=\frac{1}{2}\langle (\hat B_1 +\hat B_2 )\rangle
=sm_{-}=sm_{\perp}\exp(-\imath\varphi ) $, $m_{\perp}$ being the length of the
in-plane component of ferromagnetic vector. So, for a local condensate density
we get $n_s = s^2 m_{\perp}^2$.  It is of interest to note that in
fact all the conventional uniform $T=0$ states with nonzero $\Psi ({\bf r})$
imply simultaneous long-range order both for modulus $|\Psi ({\bf r})|$ and
phase $\varphi$. The in-plane components of antiferromagnetic vector $l_{\pm}$
describe a staggered off-diagonal order. It is worth noting that by default one
usually considers the negative sign of the transfer integral $t_{ij}$, that
implies the ferromagnetic in-plane pseudospin  ordering.

  The model exhibits many fascinating quantum phases and phase
transitions. Early investigations predict at $T=0$ charge order (CO), Bose
superfluid (BS) and mixed (CO+BS) supersolid uniform phases with an Ising-type
melting transition (CO-NO) and Kosterlitz-Thouless-type (BS-NO) phase
transitions to a non-ordered normal fluid (NO).\cite{MFA}
The detailed mean-field and spin-wave analysis of the uniform phases of 2D
hc-BH model is given by Pich and Frey.\cite{Pich}
  MFA yields for the
conventional uniform supersolid  phase \cite{Kubo}
$$
\sin^2 \beta = m_z \frac{\sqrt{V-2t}}{\sqrt{V+2t}}; \quad \sin^2 \alpha = m_z
\frac{\sqrt{V+2t}}{\sqrt{V-2t}}
$$
with a constant chemical potential $\mu = 4s\sqrt{(V^2 -4t^2)}$. It should be
noted
that the supersolid phase appears to be unstable with regard to small
perturbations in
the Hamiltonian. The mean-field energy per site of the uniform supersolid phase
is written as follows:
$$
E_{SS}=E_{CO}+s\mu m_z = E_{CO}+\mu (n_B -\frac{1}{2}),
$$
where $E_{CO}=-2Vs^2$. The cost of doping both for CO and CO+BS phase appears
to be rather high as compared with an easy-plane BS phase at half-filling where
the chemical potential turns into zero.\cite{Bernardet}

\section{Doping of BH system away from half-filling: continuous limit}
\subsection{Topological phase separation: skyrmion-like bubble domains}
Magnetic analogy allows us to make unambiguous predictions as
regards the doping of BH system away from half-filling. Indeed,
the boson/hole doping of checkerboard CO phase corresponds to the magnetization
of an antiferromagnet in $z$-direction. In the uniform easy-axis $l_z$-phase of
anisotropic antiferromagnet the local spin-flip energy cost is rather big. In
other words, the energy cost for boson/hole doping into  checkerboard CO phase
appears to be big due to a large contribution of boson-boson repulsion.
However, the  magnetization of the  anisotropic antiferromagnet in
an easy axis direction may proceed as a first order phase transition
with a ``topological phase separation'' due to the existence of
antiphase domains.\cite{AFM-domain} The antiphase domain walls
provide the natural nucleation  centers for a spin-flop phase having
enhanced magnetic susceptibility as compared with small if any
longitudinal susceptibility thus  providing the advantage of the
field energy. Namely domain walls  would specify the inhomogeneous
magnetization pattern for such an  anisotropic  easy-axis
antiferromagnet in relatively weak external magnetic field. As
concerns the domain type in quasi-two-dimensional antiferromagnet
one should emphasize the specific role played by the cylindrical, or
bubble domains which have finite energy and size. These topological
solitons have the vortex-like in-plane spin structure and often
named "skyrmions". The classical, or Belavin-Polyakov (BP) skyrmion
\cite{Belavin} describes the solutions of a non-linear
$\sigma$-model with a classical isotropic 2D Heisenberg Hamiltonian
and represents one of the generic toy model spin textures. It is of
primary importance  to note that skyrmion-like bubble domains in
easy-axis layered antiferromagnets  were actually observed  in the
experiments performed by Waldner,\cite{Waldner} that were  supported
later  by different authors (see, e.g.
Refs.\onlinecite{Kochelaev,Carsten}). Although  some questions were
not completely clarified and remain open until now,
\cite{Kamppeter,Sheka} the classical and quantum
\cite{Istomin,Kanazawa} skyrmion-like topological defects are
believed to be a genuine element of essential physics both of ferro-
and antiferromagnetic 2D easy-axis systems.

\begin{figure}[t]
\includegraphics[width=8.5cm,angle=0]{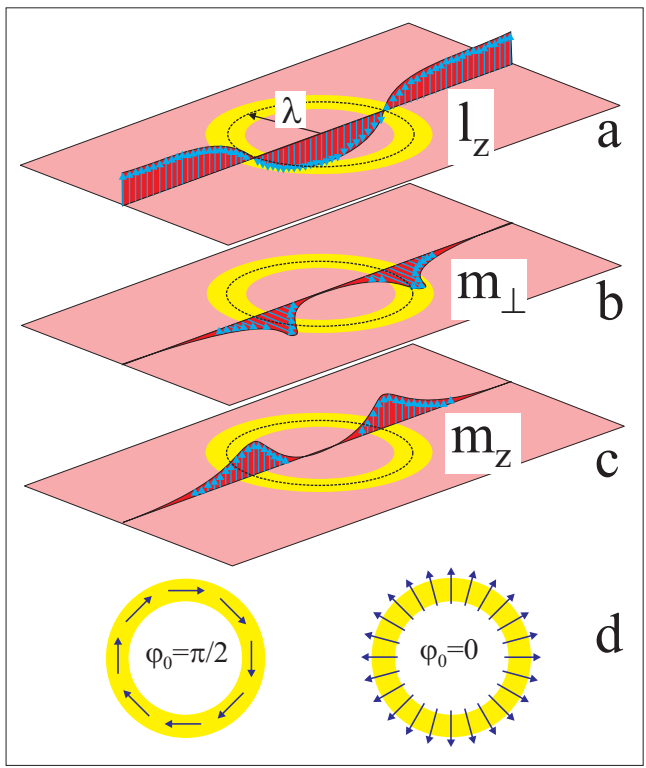}
\caption{The order parameters distribution in the skyrmion: a) The radial  distribution of the staggered charge order  parameter   $l_z$; 
b) the radial  distribution of the modulus of the superfluid order  parameter   ${\bf m}_{\perp}$; c) the radial  distribution of the charge density   ${\bf m}_{z}$ for the charged skyrmion; d) the orientation of the superfluid order  parameter   ${\bf m}_{\perp}$ given two values of the phase order parameter $\varphi _{0}$: $\varphi _{0}=0,\pi /2$, respectively. Rings in all the pictures correspond to the skyrmion (bubble domain) wall, the dashed circle  being its center.} \label{fig1}
\end{figure}

The magnetic analogy seems to be a little bit naive, however, it catches the
essential physics of doping the hc-BH system.
      As regards the checkerboard CO phase of such a  system, namely a  finite
energy  skyrmion-like  bubble domain \cite{AFM}     seems to be the  most
preferable candidate for the  domain with antiphase domain wall providing the
natural reservoir for extra bosons.
The classical description of nonlinear excitations in hc-BH model  includes the
skyrmionic solution given $V=2t$. \cite{AFM}
The skyrmion spin texture
consists of a vortex-like
arrangement of the in-plane components of ferromagnetic ${\bf m}$ vector  with
the $l_z$-component reversed
in the centre of the skyrmion and gradually increasing to match the homogeneous
background at infinity. The simplest spin distribution within classical
skyrmion
 is given as follows
$$
m_{x}=m_{\perp}\cos (\varphi + \varphi _0) ;\,
m_{y}=m_{\perp}\sin (\varphi + \varphi _0);
$$
\begin{equation}
m_{\perp}=\frac{2r\lambda }{r^{2}+\lambda ^{2}} ;
l_z=\frac{r^{2}-\lambda^{2}}{r^{2}+\lambda^{2}},
\label{sk}
\end{equation}
where $\varphi _0$ is a global phase ($U(1)$ order parameter), $\lambda$
skyrmion radius. The skyrmion spin texture  describes the
coexistence and competition of the staggered charge order parameter $l_z$ and BS order
parameter $m_{\perp}$ (see Fig.\ref{fig1}) that reflects a complex spatial interplay of potential
and kinetic  energies.
The skyrmion looks like a bubble domain in an easy-axis magnet. 
It should be noted that  the domain wall in such a
bubble domain somehow created in the checkerboard CO phase of 2D hc-BH system
represents an effective ring-shaped reservoir both for Bose-superfluid and
extra boson/hole. Indeed, the soliton   energy depends quadratically on the
 number $ \Delta n$ of bosons bound in
domain wall,\cite{AFM} similarly to that of homogeneous BS
phase.\cite{Bernardet} In other words, one might say about a zero
value of the effective boson/hole chemical potential for the CO
bubble domain configuration, provided it were a ground state. The
numerical calculations performed in frames of a classical continuous
model \cite{BH1} show that the doped bosons appear to be trapped
inside the bubble domain wall. The spatial distribution of the doped boson/hole density ($\propto m_z$) in the charged skyrmion is shown schematically in Fig. 1c. 

Skyrmions are characterized by the magnitude and sign of its topological
charge, by its size (radius), and by the global orientation of the spin. The
scale invariance of classical BP skyrmionic solution reflects in that its
energy  does not
depend on  radius, and global phase.
An interesting example of topological inhomogeneity is provided by a
multi-center
BP skyrmion \cite{Belavin} which energy does not depend on the
position of the centers. The latter are believed to be addressed
as  an additional degree of freedom, or positional order
 parameter.

In the continuous model the classical BP skyrmion is a topological excitation
and cannot
dissipate. However, the classical static skyrmion is unstable with regard to an
external field or anisotropic interactions both of easy-plane and easy-axis type.
 Small easy-axis anisotropy or external field are sufficient to shrink
skyrmion to a nanoscopic size when magnetic length $l_0$:
$$
l_0  =\left( \sqrt{\left(2V/t\right) ^{2}-\left( \mu /t\right) ^{2}}-4\right)
^{-\frac{1}{2}}
$$
is of the order of
several lattice parameters, and the continuous approximation fails to correctly
describe excitations.
Nonetheless, Abanov and Pokrovsky \cite{Abanov} have shown that the easy-axis
anisotropy together with fourth-order exchange term can stabilize skyrmion with
radius $R\propto \sqrt{l_0}.$

\subsection{Topological phase separation: Skyrmion lattices  and the low-energy
physics of BH model away from half-filling}
A skyrmionic scenario in hc-BH model allows us to make several important
predictions.
 Away from half-filling  one may
anticipate the  nucleation of a topological defect in the unconventional form
of the multi-center skyrmion-like object with  ring-shaped Bose superfluid
regions positioned in an antiphase domain wall separating the CO core
 and CO outside of the single skyrmion. The specific spatial separation of BS
and CO order parameters that avoid each other reflects the competition of
kinetic  and potential energy. Such a {\it topological} (CO+BS) {\it phase
 separation} is believed to provide a minimization of the total energy as
compared with its uniform supersolid counterpart.
Thus, the  parent checkerboard CO phase may gradually lose its stability under
boson/hole doping, while a novel topological self-organized texture is believed
to become stable.
 The most probable possibility is that every domain wall accumulates single
boson, or boson hole. Then, the number of centers in a multi-center skyrmion
nucleated with doping  has to be equal to the number of bosons/holes.   In such
a case, we anticipate the near-linear dependence of the total BS volume
fraction on the doping.
 Generally speaking, one may assume scenario
 when the nucleation of a  multi-center skyrmion  occurs spontaneously
 with no doping. In such a case we should anticipate the existence of neutral
 multi-center skyrmion-like object with equal number of positively and
negatively charged single skyrmions. However, in  practice, namely the
boson/hole doping is likely to be a physically clear driving force
for a nucleation of  a single, or multi-center skyrmion-like self-organized
collective mode in the form of multi-center charged topological defect
 which may be (not strictly correctly) referred to as
multi-skyrmion system akin in a quantum Hall ferromagnetic state of a
two-dimensional electron gas.\cite{Green} In such a case, we may characterize
an individual
skyrmion by its position (i.e., the center of skyrmionic texture), its size
(i.e., the radius of domain wall), and the orientation of the in-plane
components of pseudo-spin (U(1) degree of freedom). An isolated skyrmion is
described by the inhomogeneous distribution of the CO parameter, or staggered
boson density $l_z$, order parameter $m_z$ characterizing the deviation from
the half-filling, and $m_{\perp}$ that corresponds only to the modulus of the
superfluid order parameter.

It seems likely that for a light doping any  doped particle (boson/holes)
results in a nucleation
of a new single-skyrmion state, hence its density changes gradually with
particle doping.
Therefore, as long as the separation between skyrmionic centers is sufficiently
large so that the inter-skyrmion interaction is negligible, the energy of the
system per particle remains almost constant. This means that the chemical
potential of a boson or hole remains unchanged with doping and hence apparently
remains fixed.

 The multi-skyrmionic system in contrast with uniform ones can form the
structures
 with inhomogeneous long-range ordering of the modulus of the superfluid order
 parameter accompanied by the non-ordered global phases of single skyrmions.
Such a situation resembles in part that of granular superconductivity.

In the long-wavelength limit the off-diagonal ordering can be described by an
effective Hamiltonian in terms of  U(1) (phase) degree of freedom associated
with each skyrmion. Such a Hamiltonian
 contains a repulsive, long-range Coulomb part and a
short-range contribution related to the phase degree of freedom. The
latter term can be written out in the standard for the $XY$ model form of a
so-called Josephson coupling
\begin{equation}
H_J = -\sum_{\langle i,j\rangle}J_{ij}\cos(\varphi _{i}-\varphi _{j}),
\end{equation}
where $\varphi _{i},\varphi _{j}$ are global phases for skyrmions centered at
points $i,j$, respectively, $J_{ij}$ Josephson coupling parameter. Namely the
Josephson coupling gives rise to the long-range ordering of the phase of the
superfluid order parameter in a multi-center skyrmion. Such a Hamiltonian
represents a starting point for the analysis of disordered superconductors,
granular superconductivity, insulator-superconductor transition with $\langle
i,j\rangle$ array of superconducting islands with phases $\varphi _{i},\varphi
_{j}$. Calculating the phase-dependent part of skyrmion-skyrmion interaction
Timm {\it et al.}\cite{Timm} arrived at $negative$ sign of $J_{ij}$ that
 favors antiparallel alignment of the U(1) pseudospins. In other words,
 two skyrmions are believed to form a peculiar Josephson  $\pi$ micro-junction.
There are a number of interesting implications that follow directly from this
result:\cite{Kivelson} the spontaneous breaking of time-reversal symmetry with
non-zero supercurrents and magnetic fluxes in the ground state, long-time tails
in the dynamics of the system, unconventional Aharonov-Bohm period $hc/4e$,
negative magnetoresistence.

To account for Coulomb interaction and allow for quantum corrections we should
introduce into effective Hamiltonian  the charging energy \cite{Kivelson}
$$
H_{ch}=-\frac{1}{2}q^2 \sum_{i,j}n_{i}(C^{-1})_{ij}n_{j}\, ,
$$
where $n_{i}$ is a boson number operator for bosons bound in $i$-th skyrmion;
it
is quantum-mechanically conjugated to $\varphi$: $n_{i}=-i \partial /\partial
\varphi
_{i}$, $(C^{-1})_{ij}$ stands for  the capacitance matrix, $q$ for bosonic
charge.

The classical ground state energy of Skyrmion lattice for all
 reasonable two-dimensional lattice structures  was minimized by
 Timm {\it et al.} taking the U(1) order into account. \cite{Timm}
Besides the expected triangular lattice with frustrated antiferromagnetic U(1)
order and square lattice with N\'{e}el U(1) order the authors have also
obtained ground-state energies for all 2D Bravais lattices.
 Such a system appears to reveal a tremendously rich quantum-critical
structure. In the absence of disorder, the
$T=0$ phase diagram of the multi-skyrmion system implies either triangular, or
square crystalline arrangements (Skyrmion, or bubble crystal)
 with possible melting transition to a Skyrmion (bubble) liquid.
 \begin{figure}[t]
\includegraphics[width=8.5cm,angle=0]{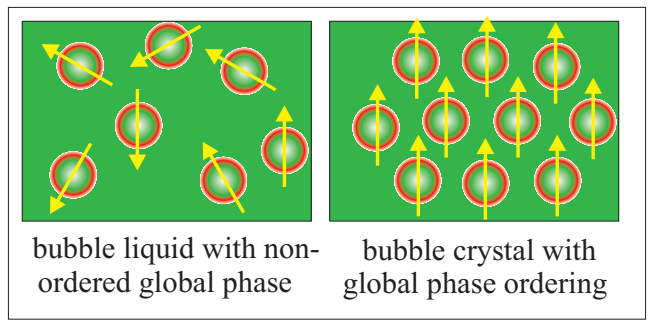}
\caption{Bubble textures for the bubble liquid  and bubble crystal. The arrows show the global phase order parameter. The left hand side panel shows snapshot of bubble texture in liquid state with non-ordered global phase. The right hand side panel illustrates the triangular bubble crystal state with "ferromagnetically" ordered global phase.
 Rings in all the pictures correspond to the bubble domain wall.} \label{fig2}
\end{figure}
The melting of Skyrmion lattice is successfully described  applying
the Berezinsky-Kosterlitz-Thouless (BKT) theory to dislocations and disclinations of the lattice and
proceeds in two steps. The first implies the transition to a
liquid-crystal phase with short-range translational order, the
second does the transition to isotropic liquid. Disorder  pins the
Skyrmion lattice and also causes the crystalline order to have a
finite correlation length. For such a system provided the skyrmion
positions  fixed at all temperatures, the long-wave-length physics
would be described by an  $XY$ model with expectable BKT transition
and gapless $XY$ spin-wave mode.

The classical phase diagram of Skyrmion lattice is quite rich. Depending on the
relation between Coulomb and Josephson coupling, and  density one may arrive at
different lattice structures with continuous or first order phase transitions.
As regards the superfluid properties the skyrmionic system reveals
unconventional behavior with two critical temperatures $T_{BS}\leq t$ and $T_c
\leq J $, $T_{BS}$ being the temperature of the ordering of the modulus and
$T_c < T_{BS}$ that of the phase of order parameter $\Psi$.
The low temperature physics in Skyrmion crystals is  governed by an
interplay of two BKT transitions,  for the U(1) phase  and
positional degrees of freedom, respectively. \cite{Timm}
Dislocations in most Skyrmion lattice types lead to a mismatch in
the U(1) degree of freedom, which makes the dislocations bind
fractional vortices and lead to a coupling of translational and
phase excitations. Both BKT temperatures either coincide (square
lattice) or the melting one is higher (triangular
lattice).\cite{Timm}
 Quantum fluctuations can substantially affect these
results. Quantum melting can destroy U(1) order at sufficiently
low densities where the Josephson coupling becomes exponentially small. Similar
situation is expected to take place in the vicinity of
structural transitions in Skyrmion crystal. With increasing the skyrmion
density
the quantum effects  result in a significant lowering of the melting
temperature as compared with classical square-root dependence.
The resulting melting temperature can reveal  an oscilating behavior as a
function of particle density with zeros at the critical (magic) densities
associated with structural phase transitions.

In terms of our model, the positional order corresponds to an incommensurate
charge density wave, while the U(1) order does to a superconductivity. In other
words, we arrive at a subtle interplay between two orders. The superconducting
state evolves from a charge order with $T_C \leq T_m$, where $T_m$ is the
temperature of a melting transition which could be termed as a temperature of
the opening of the insulating gap (pseudo-gap!?).

The normal modes of a dilute skyrmion  system (multi-center skyrmion)
include the pseudo-spin waves
propagating in-between the skyrmions, the positional fluctuations, or phonon
modes of the skyrmions which are gapless in  a pure system, but gapped  when
the lattice is pinned, and, finally,  fluctuations in the skyrmionic in-plane
orientation and  size.
The latter two types of fluctuation are intimitely
connected, since the $z$-component of
   spin and  orientation are conjugate coordinates because of  commutation
relations of quantum
   angular momentum operators. So, rotating a skyrmion changes its size.
    The orientational
   fluctuations of the multi-skyrmion system are governed by the gapless
   $XY$ model.\cite{Green} The relevant model description is most familiar as
an effective   theory of the Josephson junction array. An important feature of
the model is that it displays a quantum-critical point.

The low-energy collective excitations of skyrmion
liquid includes an usual longitudinal acoustic phonon branch.
The liquid crystal phases differ from the isotropic liquid in that they have
massive topological excitations, {\it i.e.}, the disclinations.
One should note that the liquids do not support transverse modes, these could
survive in a liquid state only as overdamped modes.  So that it is reasonable
to assume that solidification of the skyrmion lattice would be accompanied by a
stabilization of transverse modes with its sharpening below melting transition.
In other words an instability of transverse phonon modes signals the
onset of melting.

A generic property of the positionally ordered skyrmion configuration is the
sliding mode which is usually pinned by the disorder. The depinning of sliding
mode(s) can be detected in a low-frequency and low-temperature optical
response.

\subsection{Implications for the "doping-temperature" phase diagram of hc-BH
model}
\begin{figure*}[t]
\includegraphics[width=17.0cm,angle=0]{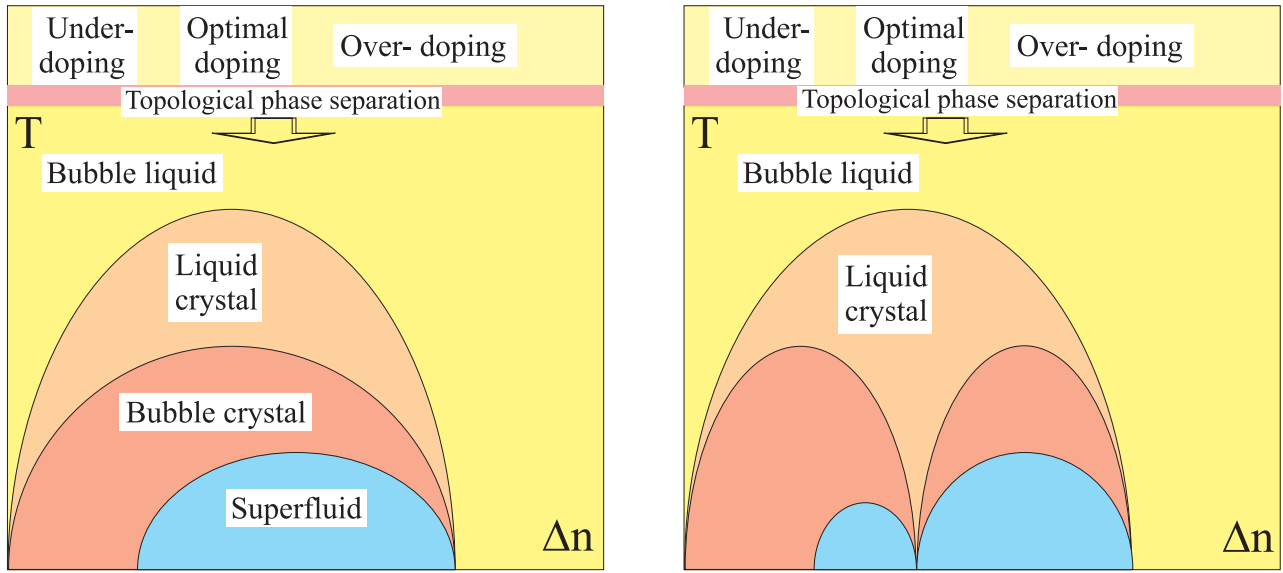}
\caption{The low-temperature part of the schematical $\Delta n$-T phase  diagram for the hc-BH model with topological phase separation.  The gradual skyrmionic solidification  evolves from the isotropic skyrmion(bubble) liquid phase, the  skyrmion liquid-crystal phase, the  incommensurate skyrmion crystal phase, and U(1) ordering, or superfluid phase, restricted by the temperatures of the proper BKT transitions.    The right hand side panel differs by
the assumed structural phase transition in the bubble lattice at a
magic doping such as $\Delta n_{b} = 1/16.$} \label{fig3}
\end{figure*}
Our  speculations as regards the topological phase separation and long
wave-length physics of 2D hc-BH model are summarized in a qualitative $\Delta
n$-T ($\Delta n_{b} = n_{b} - 0.5$) phase diagram in Fig.\ref{fig3}. First of all,  the phase diagram implies a scenario in which  the topological phase separation state evolves with a minimal doping, though it is worth noting that the ideal checkerboard CO phase  cannot   be a nominal $\Delta n_{b} = 0$ limit of any topologically phase separated phase.

Despite the qualitative character of phase diagram  we took into account some quantitative
results of quantum Monte-Carlo calculations for the 2D
hc-BH model with nearest neighbor Coulomb repulsion.\cite{Batrouni,Schmid}
First, it concerns the doping range of additive, or
near-linear concentration behavior of Bose-condensate density $\rho_s$ and  CO
structure factor $S({\bf q})$. The superfluid density increases approximately
linearly with doping
except for the most overdoped point $\Delta n_{b} \approx 0.1$, where it turns
down again.
Diagonal long-range charge order is characterized by
 the equal time structure factor at the ordering vector
${\bf q}$,
\begin{equation}
S({\bf q}) = {1 \over N} \sum_{{\bf l}}
e^{i {\bf q} \cdot {\bf l}} \langle n({\bf j},\tau)n({\bf j}+{\bf
l},\tau)\rangle.
\end{equation}
In the presence of long-range order $S({\bf q})$ will diverge with the system
size $S({\bf q})\propto L^2$ for a given ordering momentum ${\bf Q}$, which
characterizes the ordered phase. For the checkerboard order ${\bf Q}= (\pi ,\pi
)$.
In the concentration range $\Delta n_{b} = 0.0-0.1$, where both quantities
vary
linearly with $\Delta n_{b}$,\cite{Batrouni} we may approximate the topological
defect to be a system of $\Delta N =N\Delta n_{b}$ interacting
single-charged skyrmions.  It should be noted
that the both CO and BS order parameters coexist in a rather narrow doping
concentration interval: $\Delta n_{b}\leq 0.11$. Beyond this "overdoping"
region we deal with an inhomogeneous boson liquid which pseudo-spin picture
implies a strongly frustrated singlet-triplet system that resembles the spin
glass, and maybe termed as a dynamical "singlet-triplet" pseudo-spin glass.
Such a system can be characterized by a dynamical short-range   diagonal and
off-diagonal ordering with a wide distribution of respective correlation
lengths and relaxation rates. Interestingly, that in frames of a continuous
model this phase  is still described to be a strongly correlated multi-center
topological defect. However, such a model fails obviously to describe the real
system where the inter-center spacing is comparable with the lattice parameter.

The temperature evolution of hc-BH system with large inter-site boson repulsion
implies the highest critical temperature $T_{CO}(\Delta n)$  separating the
high-temperature non-ordered
NO phase (boson liquid) and a low-temperature quasi-CO phase with a disordered
system of skyrmions. The next critical temperature $T_{TPS}(\Delta n)\leq t$
points to the
 first order phase transition with a formation of
inhomogeneous Bose condensate in a single skyrmion with the
vortex-like texture of the quasi-local order parameter ${\bf
m}_{\perp}$. In other words, it is a temperature of the topological
CO+BS phase separation (TPS). In frames of our scenario such TPS
state emerges with a minimal doping, and $T_{TPS}(\Delta n_{b})$ is
likely to be nearly constant in a linear doping regime ($\Delta
n_{b}= 0.0-0.1$). The low-temperature part of the phase diagram,
which is schematically shown in Fig. \ref{fig3}, describes the
gradual skyrmionic solidification and may include the isotropic
skyrmion(bubble) liquid phase, the  skyrmion liquid-crystal phase,
the  incommensurate skyrmion crystal phase, and U(1) ordering, or
superfluid phase, restricted by the temperatures of the proper BKT
transitions. For a small deviation from half-filling
("under-doping") the temperatures of bubble crystallization/melting
are seemingly to obey the square root concentration law $T_{m}\propto \sqrt{\Delta
n_{b}}$  with a strongly developed quantum melting effect when
approaching an "overdoped" regime or concentrations  limiting the
linear regime. The superfluid phase in Fig.\ref{fig3} is arbitrarily
chosen to lie inside the  skyrmion solid phase. One should emphasize
the specific role played by quantum fluctuations: these lead to the
melting of the bubble crystal at high
 densities  and orientational disordering \cite{Timm,Rao} at low densities.
Both effects are of primary importance in the overdoped and heavily underdoped
regions of the phase diagram, respectively.
 Moreover, quantum melting effect may strongly affect the phase diagram near
the ``magic'' doping level, where the skyrmion lattice undergoes the
structural phase transition. For illustration, in the right hand
side panel in Fig. \ref{fig3} we present the possible phase diagram
for a hc-BH system with a quantum melting effect near the ``magic''
doping level $\Delta n_{b}=1/16$. In all the cases, the critical
doping level for the superfluid formation is determined by the
magnitude of the Josephson coupling constant. It is worth noting
that the bubble crystallization is accompanied by different
(pseudo)gap effects.

Our interpretation of
the phase transition at $T_{TPS}$ differs from that in
Ref.\onlinecite{Schmid}. This temperature is governed by
the magnitude  of transfer integral, and believed to be a temperature  of the
emergence of  the nonzero magnitude
of the modulus of the superconducting order parameter rather than a critical
temperature for an insulator to superconductor  transition as it is stated in
Ref.\onlinecite{Schmid}. This conclusion seems to be a result of finite size
effects and boundary conditions in QMC calculations despite the authors make
use of the most efficient QMC strategy. Such problems seem to be typical for a
finite-size simulation of many phase transitions. In addition, we should note
that MC calculations need substantial
increase in lattice size to reproduce quantitatively the long-wavelength
physics because the size of skyrmion and Skyrmion lattice parameter appear to
be   new characteristic lengths.

The close inspection of  the phase diagram in Fig.\ref{fig3}, where
$T_{c}\leq T_{m}<T_{TPS}$,  does not provide the optimistic
expectations as regards the high-temperature superconductivity in 2D
hc-BH systems, even if the magnitude of the local boson transfer
integral were as large as $t\approx 1000$ K. Nevertheless, the
attractively large temperatures $T_{TPS}$ of  the topological phase
separation with the emergence of nonzero local condensate density
engender different reasonable speculations as regards its practical
realization.

\section{Topological phase separation in discrete lattices}
The making use of the mean-field approximation   together with
simplified classical continuous models can hardly provide the
quantitative description of a quantum lattice boson system. Both
quantum effects and the discreteness of skyrmion texture can result
in  substantial deviations from the predictions of classical model.
The continuous model is relevant for discrete lattices only if we
deal with long-wave length inhomogeneities when their size is much
bigger than the lattice spacing. In the discrete lattice the very
notion of topological excitation seems to be inconsistent. At the
same time, the discreteness of the lattice itself does not prohibit
from  considering the nanoscale (pseudo)spin textures whose
topology and spin arrangement is that of a skyrmion.\cite{Gooding}
Naturally, the typical continuous models fail to describe properly
such short-wave length nanoscopic inhomogeneities. Hereafter, we
discuss a simple model which seems to catch the main effects of
discreteness and quantization.

\subsection{Nanoscopic  bubble domain in checkerboard CO phase}
What is the lattice counterpart of the  small skyrmion-like bubble domain?
\begin{figure}[h]
\includegraphics[width=8.5cm,angle=0]{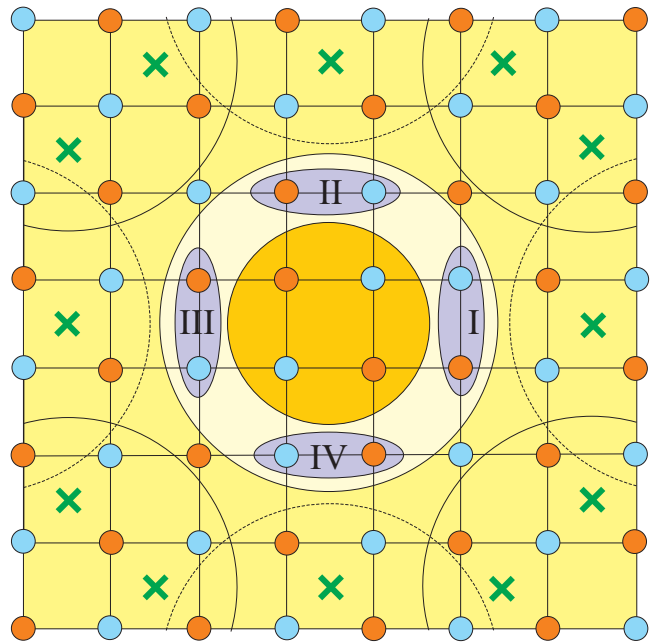}
\caption{Illustration to a skyrmion-like bubble domain in
checkerboard CO phase with 8-site ring-shaped domain wall. The four dimers within the domain wall are marked by I-IV. Schematically shown are nearest and next-nearest neighbor domains which do not overlap with the central domain. It is worth noting that there are eight additional  domains with strong nearest neighbor inter-dimer coupling with the central domain. The centers of all  12 nearest neighbor domains which do not overlap with the central domain are marked by crosses.}
\label{fig4}
\end{figure}
Fig. \ref{fig4} presents the schematic view of such a smallest
skyrmion-like bubble domain in a checkerboard CO phase for 2D square
lattice with an effective size of 4 lattice spacings. The domain
wall is believed to include as a minimum eight sites forming a
ring-shaped system of four  dimers each composed of  two sites.
There are two types of such domains which differ by a rotation by
$\pm \pi /2$. The formation of such a center seems not to require a  big
energy. Indeed, the simple estimate of a change in  potential energy:
$\Delta V \approx 4\,V_{nn} - 6\,V_{nnn}$ points to a near cancelation of $nn$ and $nnn$ contributions. The remarkable feature of the domain is in a rather small magnitude both of scalar potential and electric field inside the 8-site domain wall
(see Fig. \ref{fig4}). The flip-energy for the dimer dipole
moment is estimated to be $\delta V \approx V_{nn} - 2\,V_{nnn}$,
compared with  that for bare checkerboard CO phase ($\delta V_0
\approx 3\,V_{nn} - 4\,V_{nnn}$) that implies rather subtle
competition  between $nn$ and $nnn$ couplings. However, the difference $(\delta V_0 -\delta V )=2(V_{nn}-V_{nnn})$ is believed to be always positive and large. In other words, the flip-energy for the dimer dipole may be relatively small. Respective
dipole fluctuations would result in an effective screening of
electrostatic repulsion energy thus providing a stabilization of a
bubble-like defect. An important additional mechanism of the domain
stabilization in the hc-BH systems with local bosons composed from
electron pairs  may arise from the electron and lattice polarization
effects.\cite{Shluger,shift} As is well known, the respective
energies are comparable in value with the intersite Coulomb
interaction. Probably, namely both these effects might strongly
contribute to the  domain stabilization energy.
On the other hand, the bubble geometry implies the formation of the
electrostatic potential well inside the domain wall both for
positive and negative charges. It means that the doping into a
domain wall stabilizes the domain configuration. Such a
doping maybe    energy costless, while the energy cost of the
pseudo-spin-flip in a checkerboard CO phase is rather high: $\Delta V
\approx 4\,V_{nn}-4\,V_{nnn}$, if the neutralizing background is taken into account.

\subsection{Pseudo-spin formalism in a two-center dimer}
 With  taking into account the kinetic energy (quantum tunneling) only inside the
8-site domain wall we shall consider it as a quantum system  in an external
electrostatic potential field assuming a rigid checkerboard CO ordering overall
beyond the domain wall.  Further, taking into account the reasonable relation:
 $|t_{nn}|\gg|t_{nnn}|$, we shall consider the domain wall to be a system of
four dimers, or pairs of nearest neighboring  sites forming a quantum cluster.
\begin{figure}[t]
\includegraphics[width=8.5cm,angle=0]{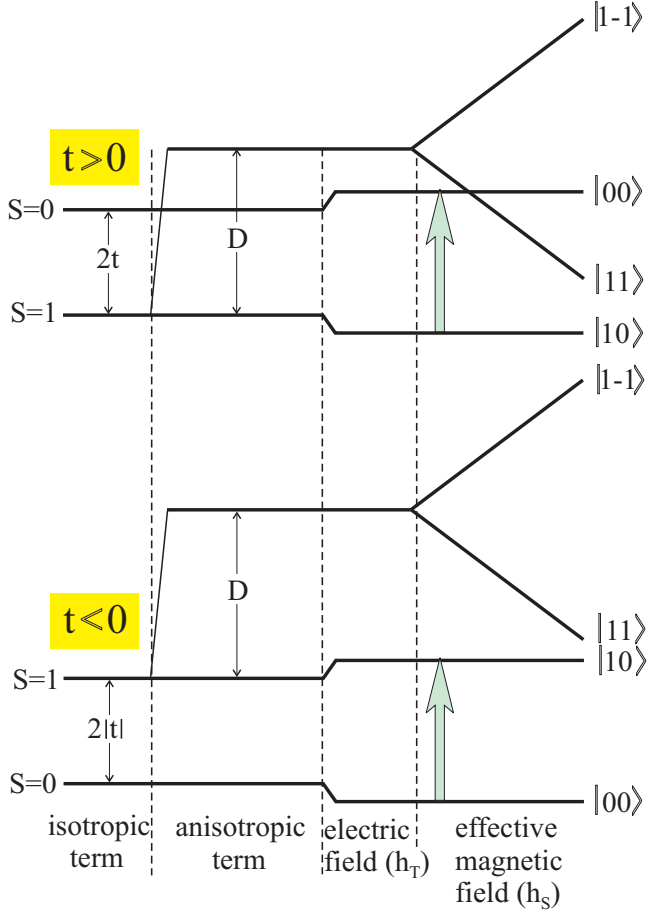}
\caption{The  step-by-step formation of the dimer energy
spectrum  given different signs of the $nn$
transfer integral. The arrows mark the
dipole-allowed transitions (see text for details).} \label{fig5}
\end{figure}
We anticipate the model though being simplified would be very instructive
analyzing the role played both by quantum and discreteness effects.

Let first address a simple two-site, or a dimer system.
The effective pseudo-spin-Hamiltonian for such a cluster, or a single dimer
Hamiltonian can be written as follows
\begin{equation}
{\hat H}_{d} = -t_{nn}[S(S+1)-\frac{3}{2}]\,+D {\hat S}_{z}^2\, -\,h_S\,{\hat
S}_{z}-\,h_T\,{\hat T}_{z}, \label{H}
\end{equation}
where ${\hat {\bf  S}}={\hat {\bf  s}}_{1}+{\hat {\bf  s}}_{2}$ is
the total pseudo-spin momentum, ${\hat {\bf  T}}={\hat {\bf
s}}_{1}-{\hat {\bf  s}}_{2}$ is an operator that changes pseudo-spin
multiplicity, $D= \frac{1}{2}V_{nn} + t_{nn}$, $h_S =\mu$, $h_T=\frac{1}{2}(V_{nn}-V_{nnn})$.
The first two terms in this effective pseudo-spin Hamiltonian describe the intra-dimer interactions, while the last two ones describe its coupling with the off-domain-wall surroundings. It is worth noting that the condition  $h_S =\mu$ means that the effective magnetic field produced by this surroundings turns into zero that provides the particle-hole symmetry of the dimer physics, in particular, for the domain
wall doping.  Actually, we arrive at an effective singlet-triplet model.\cite{STM}
Both ${\hat {\bf  S}}$ and ${\hat {\bf  T}}$ operators have a rather simple
physical sense: the former corresponds to the total "quantum" charge of the
dimer, while the latter does to the total "quantum" dipole moment. Strictly
speaking, the diagonal order parameters $\langle {\hat   S}_{z}\rangle
$ and $\langle {\hat   T}_{z}\rangle $ describe the charge and dipole
density, while the off-diagonal order parameters $\langle {\hat
S}_{\pm}\rangle $ and $\langle {\hat   T}_{\pm}\rangle $ describe the
corresponding phase ordering. It is of primary importance to note that these
order parameters are not independent because the respective operators obey the
kinematic constraint:
$$
{\hat {\bf  S}}^{2}+{\hat {\bf  T}}^{2}=3; \,({\hat {\bf  S}} \cdot {\hat {\bf
T}})=0,
$$
stemmed from a simple spin algebra.
It should be noted that there are two operators: $\hat{\bf T}={\bf s}_{1}-{\bf
s}_{2}$ and $\hat{\bf J}=[{\bf s}_{1}\times{\bf s}_{2}]$ that change the
pseudo-spin multiplicity with their matrices being symmetric and antisymmetric,
respectively:
$$
\langle 00|T_z |10\rangle =\langle 10|T_z |00\rangle =1;
$$
$$
\langle 00|J_z |10\rangle =-\langle 10|J_z |00\rangle =i \, .
$$
The Hamiltonian (\ref{H}) points to the competition of S- and T-orders in the
ring-shaped domain wall.

Fig.\ref{fig5} shows a step-by-step formation of the energy
spectrum of such a two-site  cluster, or dimer. The arrows mark the
dipole-allowed transitions that could be revealed in optical spectra.
The tunnel states $|00\rangle $ and $|10\rangle $ describing purely
"half-filled" dimer states are mixed due to the electric field with
the mixing level governed by the ratio $|h_T /2t_{nn}|$. If
$|h_{T}|\gg 2|t_{nn}|$, we arrive at classical N\'{e}el-like
"up-and-down" dimer state.

Hereafter, to describe  our singlet-triplet quantum pseudo-spin
system we start with  trial  functions
\begin{equation}
 \psi=c_0\psi_{00}+\sum_{j}(a_{j}+ib_{j}) \psi _{j}, \label{wf1}
\end{equation}
where
 the spin functions $|1M\rangle$ in Cartesian
 basis are used: $\psi_z=|10>$ and $\psi_{x,y}\sim(|11>\pm|1-1>)/\sqrt 2$.
 The conventional spin operator is represented on this basis by a simple
matrix:
 $$
 \langle\psi_i|S_j|\psi_k\rangle =-i\varepsilon_{ijk},
 $$
  and for the order parameters one
 easily obtains:
\begin{equation}
 \langle\hat{\bf S}\rangle = -2[{\bf a} \times{\bf b}],
\langle\{\hat{S}_i\hat{S}_j\}\rangle=2(\delta_{ij}-a_ia_j-b_ib_j) \label{med}
\end{equation}
given the normalization constraint $|c_{0}|^{2}+{\bf a}^2 +{\bf b}^2=1$. Thus,
for the
case of
spin-1 system the order parameters are determined by two classical vectors (two
real components of one complex vector ${\bf c} ={\bf a} +i{\bf b}$ from
(\ref{wf1})). The two vectors are coupled, so the minimal number of dynamic
variables describing the $S=1$ spin system appears to be equal to four.
\begin{figure*}[t]
\includegraphics[width=17.0cm,angle=0]{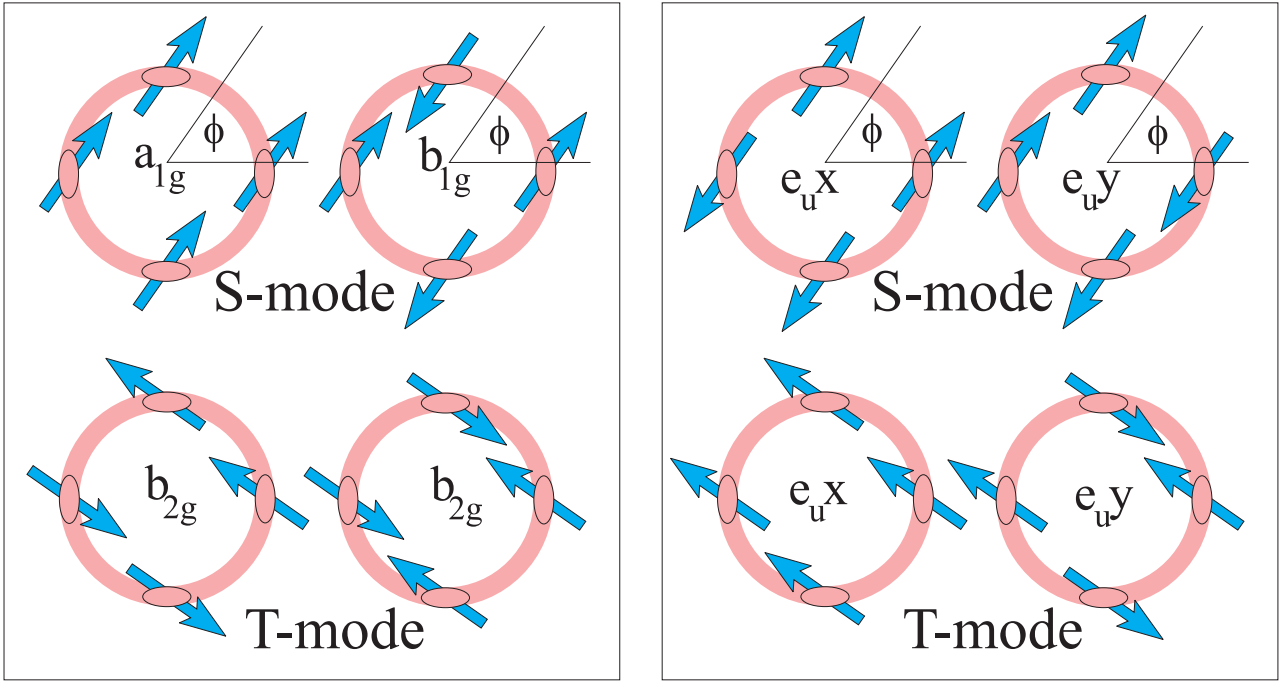}
\caption{Pseudo-spin orientation in different symmetry
superpositions for S- and T-order. Only the in-plane pseudospin
components are shown. } \label{fig6}
\end{figure*}
For the dimer as the singlet-triplet center we have additional unconventional
T,J-order parameters. For the respective averages we can easily obtain
$$
\langle\hat{\bf T}\rangle= c_{0}^{*}{\bf c}+c_{0}{\bf c}^{*}; \,
\langle\hat{\bf J}\rangle= i(c_{0}^{*}{\bf c}-c_{0}{\bf c}^{*}).
$$
Given the real value of $c_{0}$ parameter we arrive at a simple form:$\langle\hat{\bf
T}\rangle= 2c_{0}{\bf a}$, and $\langle\hat{\bf J}\rangle= -2c_{0}{\bf b}$.

\subsection{Pseudo-spin formalism and MFA description of the domain wall}

Now we may proceed with the inter-dimer coupling in the domain wall.
Introducing the symmetrized superpositions of S and T vectors, we can write
the  pseudo-spin  Hamiltonian of the inter-dimer interaction as follows
$$
{\hat H}_{dd}=\frac{1}{2}t_{nnn}[{\bf S}^{2}(a_{1g})-
{\bf S}^{2}(b_{1g}) +
({\bf S}(e_{u}){\bf T}(e_{u}))+{\bf T}^{2}(b_{2g})
$$
$$
-{\bf T}^{2}(b_{2g}^{'})]+
\frac{1}{2}(V_{nnn}-t_{nnn})[S_{z}^{2}(a_{1g})-
S_{z}^{2}(b_{1g}) +
$$
\begin{equation}
(S_{z}(e_{u})T_{z}(e_{u}))+T_{z}^{2}(b_{2g})-T_{z}^{2}(b_{2g}^{'})],
\end{equation}
where
$$
{\bf S}(a_{1g})=({\bf S}_{I}+{\bf S}_{II}+{\bf S}_{III}+{\bf S}_{IV});
$$
$$
{\bf S}(b_{1g})=({\bf S}_{I}-{\bf S}_{II}+{\bf S}_{III}-{\bf S}_{IV});
$$
$$
{\bf S}(e_{u}x)=({\bf S}_{I}+{\bf S}_{II}-{\bf S}_{III}-{\bf S}_{IV});
$$
$$
{\bf S}(e_{u}y)=(-{\bf S}_{I}+{\bf S}_{II}+{\bf S}_{III}-{\bf S}_{IV});
$$
$$
{\bf T}(b_{2g})=({\bf T}_{I}+{\bf T}_{II}-{\bf T}_{III}-{\bf T}_{IV});
$$
$$
{\bf T}(b_{2g}^{'})=({\bf T}_{I}-{\bf T}_{II}-{\bf T}_{III}+{\bf T}_{IV});
$$
$$
{\bf T}(e_{u}x)=({\bf T}_{I}+{\bf T}_{II}+{\bf T}_{III}+{\bf T}_{IV});
$$
$$
{\bf T}(e_{u}y)=({\bf T}_{I}-{\bf T}_{II}+{\bf T}_{III}-{\bf T}_{IV});
$$
(see Fig.\ref{fig4}). The chemical potential $\mu$ for the system of four dimers in a domain wall is determined now by the
condition: $\sum _{i=1}^{4}\langle S_{i}^{z}\rangle = 0,\pm 1$ for undoped and
singly-doped (boson/hole) domain wall, respectively.

The respective mean values we can address to be order parameters that describe
the subtle structure of a domain wall: as regards the diagonal order $\langle
S_{z}(a_{1g})\rangle$ specifies the full charge; $\langle S_{z}(e_{u})\rangle$
and $\langle T_{z}(e_{u})\rangle$ do the  electric dipole moments, $\langle
S_{z}(b_{1g})\rangle$ the  component of the quadrupole momentum, and one might
introduce the higher order multipole moments. As regards the off-diagonal order
we should, in general, proceed with the three types ($a_{1g},e_{u},b_{1g}$, or
$s,p,d$) of the S-order, and three types ($e_{u},b_{2g}(b_{2g}^{'})$, or $p,d$)
of the T-order, which can be defined as follows:
\begin{equation}
 \langle\hat{S}_{-}(\gamma)\rangle = \rho ^{S}_{\gamma}e^{i\varphi _{\gamma}},
 \langle\hat{T}_{-}(\gamma)\rangle = \rho ^{T}_{\gamma}e^{i\phi _{\gamma}}.
\end{equation}
Noteworthy to mention the kinematic constraint that couples different order
parameters.
In the absense of an external magnetic field the energy does not depend on
${\bf b}$ vector, it is restricted only to lie in $xy$-plane. In other words, we deal with the uncertainty of J-order parameter.

The mean-field domain-wall ground state corresponds to the coexistent ${\bf
S}(a_{1g}),{\bf T}(b_{2g})$ modes given the negative sign of the transfer
integral $t_{nnn}$ or ${\bf S}(b_{1g}),{\bf T}(b_{2g}^{'})$ modes given the
positive sign of the transfer integral $t_{nnn}$. The relation between S- and
T-mode weight is specified by the relationship between the $nn$ transfer
integral $t_{nn}$ and electric field $h_T$.
If we assume $h_T =0$, then given $t_{nn}>0$  the dimers have purely $S=1$
ground state, and we arrive at S-type off-diagonal order. In contrast,
given $t_{nn}<0$ we arrive at T-type off-diagonal order.
In general, neglecting the ST-mixing term leads to two independent phase order
parameters.
Generally speaking, the interference dipole-dipole  ST-term  would result in
ST-mixing accompanied by the constraint on the phase order parameters  with an
appearance of a noncollinearity effect.
Thus we arrive at the conclusion that the symmetry of the order parameter
distribution in the domain wall would be specified only by the sign of the
transfer integral. In addition, we see that the problem of the order parameter
associated with our bubble domain is much more complicated than in conventional
BCS-like approach \cite{sd} due to its multicomponent nature.  Moreover, we
deal with a system with different symmetry of low-lying excited states and
competing order parameters that implies their possible ambiguous manifestation
in either properties.
The low-symmetry crystalline electric fields or crystal distortions would
result in a mixing of the order parameters with different symmetry. The bubble
domain (Fig.\ref{fig4}) yields a simple and instructive toy model to describe
such effects.

It is worth noting that the flux quantization effects for the bubble domain, in
particular, the localizing effect of the magnetic field on the moving bosons,
are expected to be observed only for rather large  fields. Indeed, nanoscopic
atomic systems such as a square plaquette with a size of around 10 $\AA$
require a field of $H\approx  10^2$ Tesla for a half flux quantum per
plaquette.

Above we have concerned a simple one-center topological defect. However, in
general one has to make use of more complicated topological excitation like a
multi-center BP skyrmion.\cite{Belavin} The question arises, whether such an
entity may be described as a system of weakly coupled individual one-center
defects? In practice, it seems to be a rather reasonable  approach only for
slight deviation from half-filling, when the mean separation between doped
particles is much larger than the effective domain size.
The bubble domain may be addressed to be well isolated only if  its environment
does not contain  another domain(s) which could be involved to a "dangerous"
nearest-neighbor inter-dimer coupling.  Each domain in Fig.\ref{fig4} has
$z=12$ of such neighbors. Hence the concentration of well isolated domains can
be written as follows:
$
P_{0}(x)=x(1-x)^z,
$
where $z$ is a number of "dangerous" neighbors, $x=\Delta n_b$ the boson
concentration. The  $P_{0}(x)$ maximum  is reached at  $x_{0}=\frac{1}{z+1}$.
In our case $x_{0}=1/13$, or $\approx 0.077$. With increasing  doping the deviation of $P_{0}(x)$
from the linear law rises. On the other hand, knowing the effective domain area
$S_d \approx 9a^2$ we can roughly estimate the limiting concentration of the
single-domain model description to be $x_{max}\approx 1/9$, $\approx 0.11$. 
The inter-domain coupling includes both long-range SS- and short-range
TT-terms. We should emphasize a strong anisotropy of this coupling. For nearest
neighbors along [1,0] and [0,1] directions namely the short-range
(dipole-dipole) TT-coupling would result in an effective suppression of the
main S ordering (see Fig.\ref{fig4}), in contrast with [11] direction.

Thus we may conclude that the above mentioned simple continuous model for
Josephson junction arrays should be strongly modified to describe the
multi-domain configurations in 2D quantum hc-BH lattices, both as regards the
subtle internal domain structure, the competition of different order parameter,
and the anisotropy of Josephson coupling.

\section{Conclusions}
In conclusion, the boson/hole doping of the
hard-core boson system  away from half-filling  is assumed to be a driving
force
for a nucleation of  a  multi-center skyrmion-like self-organized
collective mode that resembles  a system of CO
bubble domains with a Bose superfluid and extra bosons both confined in domain
walls. Such a  {\it topological} CO+BS {\it phase separation}, rather than an
uniform mixed CO+BS
supersolid phase, is believed to describe the evolution of hc-BH model away
from half-filling. Starting from the classical model we predict the properties
of the respective quantum system. In frames of our scenario we may anticipate
for the hc-BH model the emergence
 of an inhomogeneous BS condensate for superhigh temperatures $T_{TPS}\leq t$,
  and 3D superconductivity for rather high temperatures $T_{c}\leq J < t$. The
system is
believed to reveal many properties typical for granular superconductors, CDW
materials,  Wigner crystals, and multi-skyrmion system akin in a quantum Hall
ferromagnetic state of a 2D electron gas.
  Topological inhomogeneity is believed  to be a generic property of 2D
  hard-core boson systems away from half-filling. Such a behavior represents a boson counterpart of the so-called {\it topological doping} being a general feature of  Mott-insulator or 2D fermion Hubbard model.\cite{Emery}

Despite all shortcomings, MFA and continuous approximation  are expected to
provide a physically clear
semiquantitative picture  of rather complex transformations taking
place in bare CO system with doping, and can be instructive as a starting point
to analyze possible scenarios. First of all, the MFA analysis allow us to
consider the  antiphase domain wall in CO phase to be a very
efficient ring-shaped potential well for the localization of a single extra
boson (hole) thus forming a novel type of a topological defect with a
single-charged domain wall. Such a defect can be addressed as a charged
skyrmion-like quasiparticle which energy can be approximated by its classical
value for CO bubble domain.
It is of great importance to note that domain wall simultaneously
represents a ring-shaped reservoir for Bose superfluid.

 Unfortunately, we have no experience to deal with multi-center
skyrmions as regards its structure, energetics, and stability.  It should be
noted that such a texture with strongly polarizable centers is believed to
provide an effective screening of
long-range boson-boson repulsion thus  resulting in an additional
self-stabilization.
Nucleation of topological phase is likely to proceed in the way typical for the
first order phase transitions.

The role played by quantum effects and  lattice discreteness has been
illustrated in frames of the simplest nanoscopic counterpart of the
bubble domain in a checkerboard CO phase of 2D hc-BH square lattice.
It is shown that the relative magnitude and symmetry of
multi-component order parameter are mainly determined by the sign of
the $nn$ and $nnn$ transfer integrals. The topologically
inhomogeneous phase of the hc-BH system away from the half-filling
can exhibit the signatures both of $s,d$, and $p$ symmetry of the
off-diagonal order. The model allows us to study the subtle
microscopic details of the order parameter distribution including
its symmetry in a real rather than momentum space, though the
problem of the structure and stability of nanoscale domain
configurations remains to be solved. The present paper establishes
only the framework for analyzing the  subtleties of the phase
separation in a lattice hc-BH model away from half-filling. Much
work remains to be done both in a macroscopic and microscopic
approaches.

I acknowledge  stimulating discussions with C. Timm, S.-L.
Drechsler, T. Mishonov, and the support by  SMWK Grant, INTAS Grant
No. 01-0654, CRDF Grant No. REC-005, RME Grant No. E 02-3.4-392 and
No. UR.01.01.062, RFBR Grant No. 04-02-96077.  I would like to thank Leibniz-Institut f\"ur Festk\"orper- und Werkstoffforschung Dresden where part of this work was made for  hospitality. Special thanks are to referees for  their remarks improving  the manuscript.


\begin{thebibliography}{99}
\bibitem{sd}
It is worth noting that in frames of BCS-like scenario the symmetry of the
order parameter is strictly defined in a momentum space albeit the discussion
of different experimental data has usually been performed with a real-space
distributed order parameter.
\end{thebibliography}
\end{document}